     \documentstyle[12pt,world_sci]{article}
\begin{document}
    \def\DZ{D\O\ }
\hyphenation{col-li-der col-li-ders cal-or-i-meter cal-or-i-meters taught}
\newcommand{\be}{\begin{equation}}
\newcommand{\ee}{\end{equation}}
\newcommand{\bdm}{\begin{displaymath}}
\newcommand{\edm}{\end{displaymath}}
\def\partder#1#2{{\partial #1\over\partial #2}}
\def\mean#1{{\mbox{$\bigl\langle #1 \bigr\rangle$}}}
\def\Mean#1{{\mbox{$\left\langle #1 \right\rangle$}}}
\def\fracerr#1{\ifmmode
                    \frac{\delta #1}{#1}
               \else
                    \mbox{${\delta #1}/{#1}$}
               \fi}

\def\D0{D\O }
\def\dedx{{\mbox{$d{\rm E}/d{\rm x}$}}}
\def\DG{D\O GEANT }
\def\GeV{{\rm GeV}}
\def\ieta{{\mbox{${\rm i}\eta$}}}
\def\iphi{{\mbox{${\rm i}\varphi$}}}
\def\MeV{{\rm MeV}}

\def\Missing#1#2{{\mbox{$#1\kern-0.57em\raise0.19ex\hbox{/}_{#2}$}}\ }

\def\vMissing#1#2{\ifmmode
            \vec{#1}\kern-0.57em\raise.19ex\hbox{/}_{#2}
         \else
            {{\mbox{$\vec{#1}\kern-0.57em\raise.19ex\hbox{/}_{#2}$}}\ }
         \fi}

\def\MEt{\Missing{E}{T}}
\def\MEx{\Missing{E}{x}}
\def\MEy{\Missing{E}{y}}
\def\MEz{\Missing{E}{z}}
\def\MExy{\Missing{E}{x,y}}
\def\trueMEt{\mbox{${\Missing{E}{T}}^{\!\!\!\!\!\!true}$}\ }
\def\vMEt{\vMissing{E}{T}}
\def\vtrueMEt{\mbox{${\vMEt}^{\!\!\!\!true}$}}

\def\St{\ifhmode {$S_T$\ }\else{S_T}\fi}

\def\sigmaMEt{\Missing{\sigma}{T}}
\def\sigmaMEx{\Missing{\sigma}{x}}
\def\sigmaMEy{\Missing{\sigma}{y}}
\def\sigmaMExy{\Missing{\sigma}{x,y}}
\def\chisq{{\mbox{$\chi^{2}$}}\ }
\def\alfas{$\alpha_{s}$ }
\def\wev{$W \rightarrow e \nu$ }
\def\Etmin{$E_{T}^{min}$ }
%
% Note that to get a "GeV." to have the correct spacing after the period, use
% "GeV\null."
% \null is shorthand for \hbox{}
%
%{\hfill D0 Note \# xxxx}
{\hfill AZPH-EXP/94-01}
     \title{Probing Color Coherency Using High pT W Events at the TEVATRON
      }
     \author{G.E. Forden\thanks{Representing the \DZ Collaboration}
     \\
     {\em Department of Physics, University Arizona, \\
     Tucson AZ}}
     %\vspace{0.3cm}
    \maketitle
     \setlength{\baselineskip}{2.6ex}

\begin{abstract}
\DZ has used $W\rightarrow e\nu$ events associated with a high pT jet to probe
for the effects of extended source color dipole radiation in W-jet rapidity
correlations.  We have also studied the low energy flow in these events
and shown an enhancement
between the jet and the beam directions, indicating the effects of color
coherency.
\end{abstract}

	Charged W bosons are produced in $p\bar p$ events predominantly
by two processes, the annihilation process and a diagram similar to
Compton scattering. In both processes a color charge is
removed from each incident hadron leaving an extended colored object, the
proton remnants.  This situation, far from being too ``dirty" to provide
useful information, is in fact a brilliant place to study radiative
corrections.
\DZ has used events where a W, observed in the $e\nu$ channel, is
produced in conjunction with a jet to study both the low energy flow
referred\cite{basic_qcd} to as the ``string drag effect" and a new measurement
designed to probe the possibility that the primary jet is radiated by a
spatially
extended colored object.  This later effect is expected to manifest itself as a
reduction of the phase space available to the jet in the large pseudo-rapidity
$(\eta)$ regions for forward W's.  We searched for this effect by studying
the W-jet rapidity correlations.

Both analyses use similar W candidate selection criteria which are
essentially those out lined in Reference 2.  In addition both
studies require the presence of a good\cite{weerts}, primary jet with some
minimum pT.  It is necessary to reconstruct the W's rapidity for both these
studies.  This is done by constraining the W candidate to have the world
average mass, 80.22 GeV.
This results, in general, in two
possible solutions for the W's longitudinal momentum.  We can not use the
electron charge to help resolve this ambiguity since \DZ does not have a
central magnetic field.  Instead, we always choose the solution with the
lower $|P_z(W)|$.  Monte Carlo studies have indicated that this gives the
correct solution the most often, given our knowledge of the event.  It should
be pointed out that W's reconstructed with large rapidities by this method
have a large ($ > 80 \% \ $) chance of being correctly reconstructed.  This
algorithm systematically moves W's which were produced with large
rapidities (but with a correspondingly smaller cross section) to the central
region, which minimizes the effect of this systematic shift.  This same
algorithm is also used on Monte Carlo events in any comparisons
performed.

The lowest order diagrams for the production of a high pT W all have an
internal quark line whose virtuality is directly related to the W-Jet system
invariant mass.  This mass tends to be minimized by the internal quark
propagator.  This can be accomplished when the W and jet have the same
rapidity, but of course are opposite in $\phi$.  This behavior can be
modified by various processes.  The parton density functions might
systematically inhibit various kinematicly allowed topologies.  Color
coherence effects producing radiated gluons preferentially between the
primary jet and the beam remnants could systematically recoil the primary jet
towards lower rapidities.  Another possibility, as mentioned above, is that
interference effects involving all the incident hadrons\cite{Bo} constituents
could restrict the primary jet to central rapidities.

We have studied the W-jet rapidity correlation for events where there is a jet
produced with $pT > 16 GeV$ and compared it to various Monte Carlos
reflecting lowest order\cite{papa}, next to leading order\cite{dyrad} and
extended color dipole\cite{lonnblad} predictions.  The average jet rapidity
as a function of the W rapidity is shown in Figure 1 together with the
predictions for the various Monte Carlos.  It is clear that none of the Monte
Carlos do a particularly good job of describing the data.  However, by
performing a $\chi^2$ analysis on the actual jet rapidity distributions for
those events with W rapidities greater than 1.0 we
can exclude the leading order and the next to leading order at the 95\%
confidence level while there is an 18\% chance that the discrepancies
between ARIADNE, the color dipole model, and the data are due to
statistical fluctuations alone.

%}
%\caption{ The average jet rapidity of the primary jet as a function of
%%reconstructed
% W rapidity.  The LO, NLO and ARIADNE Monte Carlo predictions are also shown.
%}

We select those $W\rightarrow e\nu$ events which have a central jet with pT $>$
10 GeV and a reconstructed central W to compare the energy flow in an
annulus around the jet and the corresponding annulus around the W.  Both
production mechanisms, the annihilation and the Compton scattering,
produce at least one color string between the jet and a beam remnant (the
annihilation process produces two, one to each remnant) and none to the
color neutral W.  The string drag effect should produce an increased energy
flow on the line between the jet and either beam while this increase should
not be present on the W side of the event.
The inner radius
of each annulus is $\Delta R=0.7$ which corresponds to the standard \DZ
jet cone while the outer radius is $\Delta R=1.5$.
Polar coordinates
centered on these cones are used to determine the energy flow with respect to
the closest beam direction.  We define a polar angle $\beta$ which is zero in
the
direction of the beam.  The energy flow  $dE/d\beta$ is then the radial
integral of this energy as a function of $\beta$.  The resulting measurement
of the energy around the jet is shown in Figure 2.  The corresponding
energy around the W does not show this enhancement at $\beta=0$, the
direction toward the beam.  This is in qualitative agreement with analytic
calculations by Dokshitzer et al. \cite{dok}

%\caption{
%    The energy flow around the jet, $dE/d\beta$, as a function of the angle
%%$\beta$
%    which is zero in the direction of the beam.
%}

\DZ has measured the rapidity correlation between high pT W's and the
primary jets.  We find that the jet stays central almost independent of the
W's rapidity.  This is in conflict with the leading order (LO) and next to
leading order (NLO) QCD expectations which predict that the jet should
follow the W's rapidity.  The Monte Carlo ARIADNE, which is based on
radiation patterns being determined by extended color dipoles, qualitatively
predicts this and can not be ruled out.  We have
also shown that the low energy flow in high pT W events is in qualitative
agreement with
coherent radiation pattern predictions.

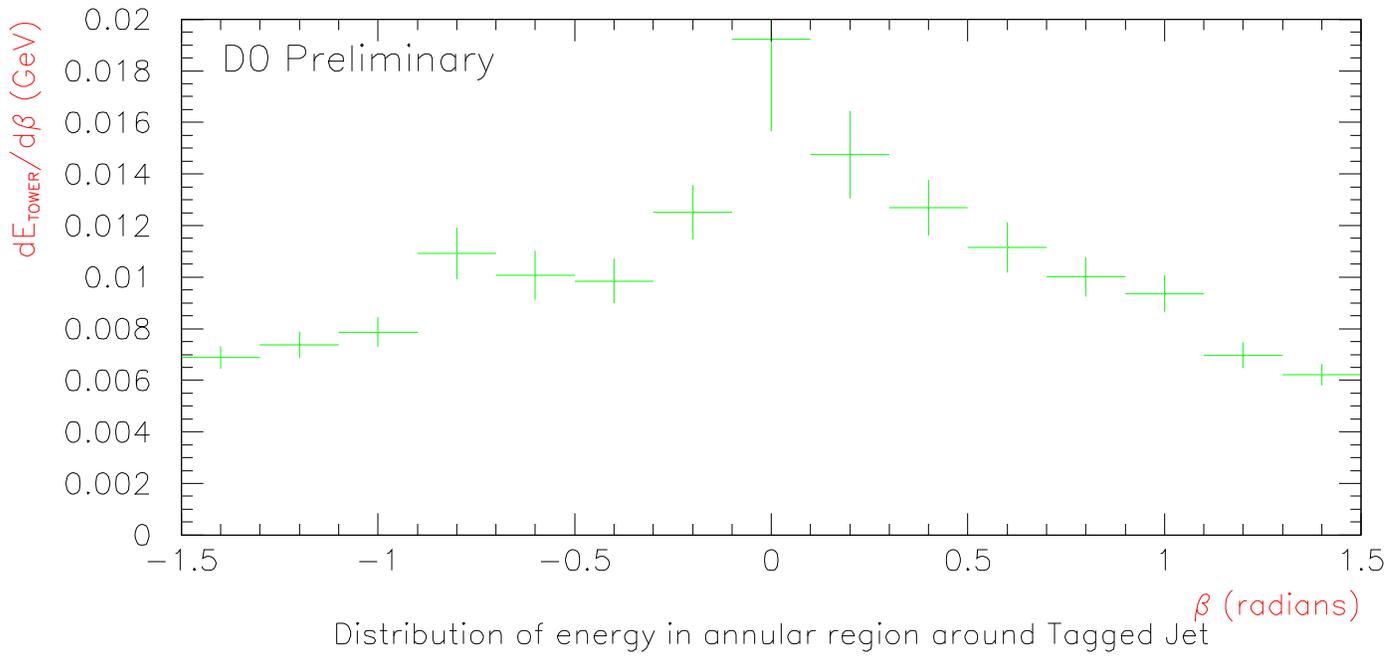
\end{document}